\documentstyle[preprint,aps]{revtex}
\begin{document}

\draft

\title
{\bf Binding energy of exciton complexes determined by the
tunnelling current of single electron transistor under optical
pumping}

\author{ David M.-T. Kuo }
\address{Department of Electrical Engineering, National Central
University, Chung-Li, Taiwan, Republic of China}

\author {Yia-Chung Chang}
\address
{Department of Physics\\
University of Illinois at Urbana-Champaign, Urbana, Illinois
61801}


\date{\today}
\maketitle
\begin{abstract}
We theoretically study the tunnelling current of a single electron
transistor (SET) under optical pumping. It found that holes in the
quantum dot(QD) created by optical pumping lead to new channels
for the electrons tunnelling from emitter to collector.  As a
consequence, an electron can tunnel through the QD via additional
channels, characterized by the exciton, trion and biexciton
states. The binding energy of exciton complexes can be determined
by the Coulomb oscillatory tunnelling current.
\end{abstract}

\newpage

Recently, the spontaneous emission spectrum of a single quantum
dot (QD) has been suggested as a single-photon source, which is
important in the application of  quantum cryptography$^{1-4}$.
Experimentally, the spontaneous emission spectrum typically
exhibits coexisting sharp emission peaks, which have been
attributed to the electron-hole recombination in the exciton,
trion, and biexciton formed in the QD$^{5}$. Nevertheless,it is
difficult to experimentally determine the binding energy of
exciton complexes due to lack of free electron-hole recombination
in the QD. In this letter we propose that the tunnelling current
of single electron transistor (SET) under optical pumping can be
employed to determine the binding energy of exciton. Our
calculation is based on the Keldysh Green's function approach
within the Anderson model for a two-level system. We find that the
optical excitation creates holes in the QD, which provide new
channels (via the electron-hole interaction) for the electron to
tunnel from the emitter to the collector. Consequently, an
electron can tunnel through the QD via four additional channels,
characterized by the exciton, positive trion, negative trion, and
biexciton states. Each addition channel can generate a new
oscillatory peak in the tunnelling current characteristics in
addition to the typical peaks caused by the electron-electron
Coulomb interactions. The binding energy of exciton complexes as
well as electron charging energy can be determined by using the
tunnelling current as functions of gate voltage.

The system under the current study considers a single quantum dot
(QD) sandwiched between two leads. Electrons are allowed to tunnel
from the left lead (emitter) to the right lead (collector) under
the influence of an optical pump. We start with the following
Hamiltonian
\begin{equation}
H=H_d+H_l+H_{d,l}+H_{d,e}+H_I,
\end{equation}
where the first term describes electrons in the InAs/GaAs QD . We
assume that the quantum confinement effect is strong for the small
QD considered here. Therefore, the energy spacings between the
ground state and the first excited state for electrons and holes,
$\Delta E_{e}$ and $\Delta E_{h}$, are much larger than thermal
energy, $k_B T$, where $k_B$ and $T $ denote the Boltzmann
constant and temperature. Only the ground state levels for
electrons and holes, $E_e$ and $E_h$, are considered in $H_d$. The
second term describes the kinetic energies of free electrons in
the electrodes, where the correlation effects among electrons is
ignored. Note that in the current setup, the gate electrode does
not provide any electrons, but merely controls the energy levels
of the QD. The third term describes the coupling between the QD
and the leads. The fourth term describes the interband optical
pumping with a frequency of $\omega_0$, which is in resonance with
the energy difference between an electron level in the wetting
layer and the hole ground state level. Due to the large
strain-induced splitting between the heavy-hole and light-hole
band for typical QDs, we only have to consider the heavy hole band
(with $J_z=\pm 3/2$) and ignore its coupling with light-hole band
caused by the QD potential. Because the effect of inter-particle
Coulomb interactions is significant in small semiconductor QDs, we
take into account the electron Coulomb interactions and
electron-hole Coulomb interactions in the last term.

Once the Hamiltonian is constructed, the tunnelling current of SET
can be calculated via the Keldysh Green's function method.$^{6}$
We obtain the tunnelling current through a single dot
\begin{equation}
J=\frac{-2e}{\hbar}\int \frac{d\epsilon}{2\pi}[f_L(\epsilon-\mu_L)
-f_R(\epsilon-\mu_R)] \frac{\Gamma_L(\epsilon) \Gamma_R(\epsilon)}
{\Gamma_L(\epsilon)+\Gamma_R(\epsilon)}ImG^r_{e,\sigma}(\epsilon).
\end{equation}
Eq.~(2) is still valid for the SET under optical pumping provided
that the condition $\Gamma_{L(R)}>> R_{eh}$ is satisfied, where
$R_{eh}$ is the electron-hole recombination rate. $f_L(\epsilon)$
and $f_R(\epsilon)$ are the Fermi distribution function for the
source and drain electrodes, respectively. The chemical potential
difference between these two electrodes is related to the applied
bias via $\mu_L-\mu_R=eV_a$. $\Gamma_L(\epsilon)$ and
$\Gamma_R(\epsilon)$ denote the tunnelling rates from the QD to
the left (source) and right (drain) electrodes, respectively. For
simplicity, these tunnelling rates will be assumed energy and
bias-independent. Therefore, the calculation of tunnelling current
is entirely determined by the spectral function
$A=ImG^r_{e,\sigma}(\epsilon)$, which is the imaginary part of the
retarded Green's function $G^r_{e,\sigma}(\epsilon)$.

The expression of retarded Green's function
,$G^{r}_{e,\sigma}(\epsilon)$, can be obtained by the equation of
motion of $G^r_{e,\sigma}(t)=-i\theta(t)\langle
\{d_{e,\sigma}(t),d^{\dagger}_{e,\sigma}(0)\}\rangle $, where
$\theta(t)$ is a step function, the curly brackets denote the
anti-commutator, and the bracket  $\langle ...\rangle  $
represents the thermal average. After some algebras$^{5}$, the
retarded Green's function of Eq.~(2)
\begin{small}
\begin{eqnarray}
&  & G^r_{e,\sigma}(\epsilon)=
(1-N_{e,-\sigma})\{\frac{1-(n_{h,\sigma}+n_{h,-\sigma})+n_{h,\sigma}n_{h,-\sigma}}
{\epsilon - E_{e} +i \frac{\Gamma_{e}}{2} }\\ \nonumber &
+&\frac{n_{h,\sigma}+n_{h,-\sigma}-2
n_{h,\sigma}n_{h,-\sigma}}{\epsilon - E_{e} + U_{eh} +
i\frac{\Gamma_e}{2}} +\frac{n_{h,\sigma}n_{h,-\sigma} }{\epsilon -
E_{e} + 2U_{eh}+ i\frac{\Gamma_{e}}{2}}\}\\ \nonumber& + &
N_{e,-\sigma}\{\frac{1-(n_{h,\sigma}+n_{h,-\sigma})+n_{h,\sigma}n_{h,-\sigma}}
{\epsilon - E_{e}- U_{e} + i\frac{\Gamma_{e}}{2}}\\ \nonumber&+&
\frac{n_{h,\sigma}+n_{h,-\sigma}-2
n_{h,\sigma}n_{h,-\sigma}}{\epsilon- E_{e}- U_{e}+U_{eh} +
i\frac{\Gamma_{e}}{2}}+ \frac{n_{h,\sigma}n_{h,-\sigma}}{\epsilon
- E_{e}- U_{e}+ 2U_{eh}+ i\frac{\Gamma_{e}}{2}} \}.
\end{eqnarray}
\end{small}

In Eq.~(3), $\Gamma_e$ is the electron tunnelling rate $\Gamma_e
\equiv \Gamma_L+\Gamma_R$. It is worth noting that the
electron-hole recombination effect have not been directly included
into $G^{r}_{e,\sigma}(\epsilon)$, because we assumed that
$\Gamma_e
>> R_{eh}$. The typical value of $R_{eh}$ for InAs/GaAs QDs is $\sim
1/ns$. The electron occupation number of the QD can be solved in a
self-consistently via the relation

\begin{small}
\begin{equation}
N_{e,\sigma} = -\int \frac{d\epsilon}{\pi} \frac{\Gamma_L
f_L(\epsilon)+\Gamma_R f_R(\epsilon)}{\Gamma_L+\Gamma_R}
ImG^r_{e,\sigma}(\epsilon).
\end{equation}
\end{small}
$N_{e,\sigma}$ is limited to the region $0 \le N_{e,\sigma} \le 1
$. Eq.~(4) indicates that the electron occupation numbers of the
QD, $N_{e,-\sigma}$ and $N_{e,\sigma}$, are primarily determined
by the tunnelling process. To obtain the electron and hole
occupation numbers ($n_{e,-\sigma}=n_{e,\sigma}$ and
$n_{h,-\sigma}=n_{h,\sigma}$) arised  from the optical pumping, we
solve the rate equations and obtain
\begin{equation}
n_e=n_{e,-\sigma}=n_{e,\sigma}=\frac{\gamma_{e,c} N_{e,k}}
{\gamma_{e,c} N_{e,k} + R_{eh} n_{h}+ \Gamma_e},
\end{equation}
and
\begin{equation}
n_h=n_{h,-\sigma}=n_{h,\sigma}=\frac{\gamma_{h,c} N_{h,k}}
{\gamma_{h,c} N_{h,k} + R_{eh}(n_e + N_e)+ \Gamma_h },
\end{equation}
where $\gamma_{e(h),c}$ and $N_{e(h),k}$ denote the captured rate
for electrons (holes) from the wetting layer to the QD and the
occupation number of electrons (holes) in the wetting layer. Here,
we assume that $N_{e(h),k}$ is in proportion to the intensity of
excitation power, $ p_{exc}$. $\Gamma_h$ denotes the nonradiative
recombination rate for holes in the QD.

According to Eq.~(3), particle Coulomb interactions will
significantly affect the tunnelling current of SET. To illustrate
this effect, we apply our theory to a self-assembled InAs/GaAs QD
with pyramidal shape. First, we calculate the inter-particle
Coulomb interactions using a simple but realistic effective-mass
model. The electron (hole)in the QD is described by the equation

\begin{eqnarray}
&& [-\nabla \frac {\hbar^2} {2m_{e(h)}^*(\rho,z)} \nabla +
V^{e(h)}_{QD}(\rho,z)\mp eFz] \psi_{e(h)}({\bf r}) \nonumber \\ &=
& E_{e(h)} \psi_{e(h)}({\bf r}),
\end{eqnarray}
where ${m_e^*(\rho,z)}$ (a scalar) denotes the position-dependent
electron effective mass, which has $m_{eG}^* = 0.067 m_e$ for GaAs
and $m_{eI}^* = 0.04 m_e$ for InAs QD. ${m_h^*(\rho,z)}$ denotes
the position-dependent effective mass tensor for the hole. It is a
fairly good approximation to describe ${m_h^*(\rho,z)}$ in
InAs/GaAs QD as a diagonal tensor with the $x$ and $y$ components
given by ${m^*_t}^{-1}=(\gamma_1+\gamma_2)/m_e$ and the $z$
component given by ${m^*_l}^{-1}=(\gamma_1-2\gamma_2)/m_e$.
$\gamma_1$ and $\gamma_2$ are the Luttinger parameters.
$V^e_{QD}(\rho,z)$ ($V^h_{QD}(\rho,z)$) is approximated by a
constant potential in the InAs region with value determined by the
conduction-band (valence-band) offset and the deformation
potential shift caused by the biaxial strain in the QD. These
values have been determined by comparison with results obtained
from a microscopic model calculation$^{7}$ and we have $V^e_{QD}=
-0.5 eV$ and $V^h_{QD}=-0.32 eV$. The $eFz$ term in Eq.(7) arises
from the applied voltage, where $F$ denotes the strength of the
electric field. Using the eigenfunctions of Eq.~(7), we calculate
the inter-particle Coulomb interactions via
\begin{equation}
 U_{i,j}= \int d{\bf r}_1 \int d{\bf r}_2 \frac {e^2 [n_i({\bf r}_1)n_j({\bf r}_2)]}
{\epsilon_0 |{\bf r}_{1}-{\bf r}_{2}|},
\end{equation}
where $i(j)=e,h$. $n_i({\bf r}_1)$ denotes the charge density.
$\epsilon_0$ is the static dielectric constant of InAs.  The
Coulomb energies are different in different exciton complexes, but
the difference is small.$^{5}$ Therefore, only the direct Coulomb
interactions have been taken into account in this study.

For the purpose of constructing the approximate wave functions, we
place the system in a large confining cubic box with length L.
Here we adopt $L=40 nm$. The wave functions are expanded in a set
of basis functions, which are chosen as sine waves
\begin{small}
\begin{eqnarray}
\psi_{nlm}(\rho,\phi,z)&= &\frac{\sqrt{8}}{\sqrt{L^3}} sin (k_lx)
sin (k_m y)sin (k_n z) ,
\end{eqnarray}
\end{small}
where $k_n = n\pi/L$,$k_m = m\pi/L$,$k_\ell = \ell \pi/L$. n, m
and $\ell$ are positive integers. The expression of the matrix
elements of the Hamiltonian of Eq.~(7) can be readily obtained. In
our calculation $n=20$, $m=10$ and $\ell=10$ are used to
diagonalize the Hamiltonian of Eq.~(7). Fig. 1 shows the
inter-particle interactions as functions of QD size. The ratio of
height and base length is $h/b=1/4$, while $h$ varies from $2.5nm$
to $6.5 nm$. The strengths of Coulomb interactions are inversely
proportional to the QD size. However as the QD size decreases
below a threshold value (around $b=12 nm$), $U_{e}$ is
significantly reduced due to the leak out of electron density for
small QDs. These Coulomb interactions approach approximately the
same value in the large QD limit. This indicates similar degree of
localization for electron and hole in large QDs. We also note that
$U_{eh}$ is smaller than $U_{e}$ in the large QD. This is due to
the fact that in large QDs the degree of localization for the hole
becomes similar to that for electron, while the anisotropic nature
of hole wave function reduces $U_{eh}$. The repulsive Coulomb
interactions, $U_{e}$ and $U_{h}$, are the origin of Coulomb
blockade for electrons and holes, respectively. The attractive
Coulomb interaction $U_{eh}$ gives rise to the binding of the
exciton. To study the behavior of tunnelling current, we consider
a particular pyramidal InAs/GaAs QD with base length $b=13$nm and
height $h=3.5nm$. The other relevant parameters for this QD are
$E_{e,0}=-0.14 eV$, $E_{h,0}=-0.125 eV$,$U_{e}=16.1meV$,
$U_{eh}=16.7 meV$ and $U_{h}= 18.5 meV$.

Now, we perform detailed numerical calculation of the tunnelling
current. For simplicity, we assume that the tunnelling rate
$\Gamma_L = \Gamma_R = 0.5 meV$ is bias-independent. We apply a
bias voltage $V_a$ across the source-drain and $V_g$ across the
gate-drain. The QD electron and hole energy levels, $E_e$ and
$E_h$, will be changed to $E_e+\alpha eV_a - \beta eV_g$ and
$E_h+\alpha eV_a - \beta eV_g$, where $\alpha$ and $\beta$ are the
modulation factors. In our calculation, we assume $\alpha =0.5 $
and $\beta = 0.7$, which can also be determined by
experiments$^{9}$. Meanwhile the chemical potentials of the
electrodes with Fermi energy $E_F=60 meV$ (relative to the
conduction band minimum in the leads) are assumed to be $70 meV$
below the energy level of $E_{e}$ at zero bias. Parameters
$\Gamma_h=0.2 meV$ and $R_{eh} = 10 \mu eV$ are adopted.

Applying Eqs.~(2), (4) and (6), we solve for the electron
occupation number $N_e=N_{e,\sigma}=N_{e,-\sigma}$ and tunnelling
current $J$. Fig. 2 shows the calculated results for $N_e$ and $J$
as functions of gate voltage with and without the
photon-excitation power at zero temperature and $V_a= 2 mV$. Solid
line and dashed line correspond to $I=0$ (no pump) and $I=0.9$
(with pump), respectively. We have defined a dimensionless
quantity, $I\equiv \gamma_{h,c} N_{h,k}/\Gamma_h$, which is
proportional to the pump power. The electron occupation number
displays several plateaus, while the tunnelling current displays
an oscillatory behavior. We label four critical voltages (from
$V_{g1}$ to $V_{g4}$) to indicate the resonance energies of
retarded Green's function. We see that the photon-excitation leads
to additional two current peaks below the voltage $V_{g3}$, which
is caused by the electron tunnelling assisted by the presence of a
hole in the QD. This interesting phenomenon was observed by
Fujiwara et al. in an SET composed of one silicon (Si) QD and
three electrodes$^{9}$. The behavior of the photo-induced
tunnelling current can be understood by the analysis of the poles
of retarded Green's function of Eq.~(3); the first peak of dashed
line corresponds to the tunnelling current through the energy
level at $\epsilon=E_e-2 U_{eh}$ (corresponding to a positive
trion state) . The second peak is caused by a pair of poles at
$\epsilon=E_e - U_{eh}$ (the exciton state) and
$\epsilon=E_e+U_e-2U_{eh}$ (the biexciton state). Since the
magnitude of $U_e$ is very close to that of $U_{eh}$, these two
poles almost merge together. The third peak is caused by another
pair of poles at $\epsilon=E_e$ (the single-electron state) and
$\epsilon=E_e+U_e -U_{eh}$ (the negative trion state). The last
peak locating near $V_{g}=123 mV$ is due to the tunnelling current
through the energy level at the pole $\epsilon=E_e+U_e$ (the
two-electron state). The gate voltage difference $\Delta
V_{g21}=V_{g2}-V_{g1}$ $(V_{g43}=V_{g4}-V_{g3})$ determine the
strength of electron-hole interaction $\beta \Delta V_{g21}
=U_{eh}$ ($\beta V_{g43}=U_e$). Once $U_{eh}$ is determined, we
obtain the binding energy of exciton complexes.

In this study we have used the tunnelling current of an SET under
optical pumping to determine the electron-hole interaction
$U_{eh}$, which can be regarded as the binding energy of exciton.
Although we used InAs/GaAs SET as an example, this idea can also
be applied to $Si/SiO_2$ SET system$^{9}$.

{\bf ACKNOWLEDGMENTS}

This work was supported by National Science Council of Republic of
China under Contract Nos. NSC 93-2215-E-008-014 and
NSC-93-2120-M-008-002


\mbox{}

\newpage

{\bf Figure Captions}

Fig. 1: Intralevel Coulomb interactions $U_{e}$ and $U_{h}$ and
interlevel Coulomb interaction $U_{eh}$ as a function of the QD
base length $b$.

Fig. 2: Electron occupation number $N_e$ and tunnelling current as
functions of gate voltage at zero temperature for various
strengths of optical excitation. Current density is in units of
$J_0=2e \times meV/h$.

\end{document}